\documentstyle[12pt]{ichep}
\begin{document}
\title{Beyond the Standard Model }
\author{S.\ Dimopoulos$^{\dag}$ }
\affil{CERN, Geneva, Switzerland$^{\S}$ }
\abstract{A few topics beyond the standard model are reviewed. }
\twocolumn[\maketitle]
\fnm{1}{This document was prepared by I.G.\ Knowles based on a
recording of the plenary talk given at the conference.}
\fnm{3}{On leave from Stanford University, Stanford, CA 94305, USA.
\\
Work supported in part by NSF grant, NSF-PHY-9219345. }

\noindent
\section{INTRODUCTION}
The subject of physics beyond the Standard Model (SM) began
flourishing
around the year 1978. In early 1978 SLAC discovered parity violation
in
neutral processes. That convinced many ambitious theorists
that the SM was correct and they started to focus on the next layer
of
fundamental questions. At this point a dichotomy started emerging
between
theory and experiment. Theorists began focusing on speculative
ideas.
These came in basically four categories:
\begin{itemize}
\item Unification [1,2];
\item Technicolor \cite{tech};
\item Supersymmetric (SUSY) Unification [4,5];
\item Superstrings \cite{string}.
\end{itemize}
and they concentrated mostly on the blemishes of the SM and on
reasons
why it cannot be a fundamental theory. On the other hand our
experimentalist
friends have been confirming the SM year after year. $W^\pm\;Z^0$
have
been discovered at CERN [7,8] and the top at Fermilab \cite{disct}
and high
level precision electroweak tests by LEP \cite{prec} have been
vindicating
the SM over and over again. So, to first approximation it is fair to
say
that there is no need to go beyond the SM and that therefore this
talk is
unnecessary.

This is indeed the situation to first approximation, except for a
small,
but perhaps significant, hint that has emerged recently. This hint
comes
from the weak mixing angle measured by the LEP experiments
\cite{hint1}
and SLD \cite{hint2}. Supersymmetric unified theories
that were proposed in 1981 [4,5] predicted the weak mixing angle,
$\sin^2
\theta_W$ to within a theoretical uncertainty of approximately
$\pm$1\%.
Recent experimental measurements have measured this angle to roughly
$\pm
0.2$\%. The theoretical prediction agrees very
well with experiment. Now of course this could just be a coincidence;
the {\it a priori} probability for this is 2\%. If you adopt the
viewpoint that this is just a coincidence then you really have no
hint of
physics beyond the SM. We will adopt a different viewpoint, we will
take
this coincidence seriously and we will pursue the consequences of it.

I should remark that we are not at all alone in taking this
coincidence
seriously. Of the 95 abstracts and 66 papers that were submitted to
this
session more than
three quarters dealt with supersymmetric unified theories
\cite{subs}.
Also if you look at the hep-ph phenomenology bulletin board
\cite{bull}
you will notice that roughly a quarter of all papers that are
submitted
deal with SUSY So a Martian that just looks at the titles of hep-ph
phenomenology might be confused as to whether SUSY has or has not
been
found.

Of course I do not have time to cover all the contributions to these
proceedings [15,16,17] in detail but I will occasionally refer
to some of the results that these people have reported.

My talk consists of three parts. First I will discuss the question of
the
weak mixing angle in SUSY Grand Unified Theories (GUTs) and in
general the
question of why SUSY GUTs were proposed and what are some of their
virtues.
Then I will discuss the top, how it fits in SUSY
GUTs and how it may fit in the SM. Finally I will make some brief
remarks
about theories that attempt to make statements about the masses and
mixing angles of other quarks and leptons. I will not have time to
review
technicolor which has already been discussed in some detail by K.\
Lane
\cite{lane}.

\section{WHY SUSY GUTs}

Let me begin by reminding you very briefly why SUSY GUTs were
proposed and what are some of their virtues. We begin with the
fundamental
premise that theorists believe in, that there is a fundamental
scale in nature. This is near the gravitational scale, the Planck
mass
$M_{Pl}$, of the order $10^{18}$ GeV$/c^2$. An important question
before even beginning to do physics is: can we discuss physics at
our energies without knowing almost anything about the physics at
this
fundamental scale?\\

\subsection{The Decoupling Hypothesis}
The basic hypothesis that allows us to begin and go forward is the so
called decoupling hypothesis. It says
that the answer to the above question is yes. This hypothesis is very
intuitive: it is the same reason for example that in cooking
schools they don't teach you nuclear physics.
It allows us to discuss large distance physics while being
ignorant about what happens at short distances.

The quantitative statement of the decoupling hypothesis is
that low energy physics parameters are fairly insensitive to the
Planck
mass or to this fundamental scale. They do not depend on positive
powers
of this scale, $M^n_{Pl}$, they are at most logarithmic functions of
the
Planck mass, $\log M_{PL}$.

Now the vast number of theories violate this decoupling hypothesis.
The
first class of such theories, which covers almost all theories, are
the
so-called non-renormalizable theories. These are maximally
non-decoupled:
in order to parameterize your ignorance of Planck scale physics you
need
infinitely many parameters all of which depend on positive powers of
$M_{Pl}$. The second class of theories, like the SM, are the
renormalizable
theories. In these theories you can parameterize your ignorance
of the fundamental physics with a few parameters, ${\cal O}(20)$,
most of
parameter, the Higgs mass (or scalar masses in general), which is
actually
very sensitively dependent on details of Planckian physics. Finally
there are SUSY theories which are totally decoupled in the sense
that all parameters depend at most logarithmically on $M_{Pl}$.

Now I should emphasize that this logarithmic dependence on $M_{Pl}$
is
actually very important. The weak mixing angle actually depends
logarithmically on $M_{Pl}$ and the experimental measurement that
determines
it is an indirect measurement of physics at the Planck scale
(actually the
unification scale).

The decoupling hypothesis was the original reason why SUSY GUTs were
proposed. In order for SUSY to help you totally decouple low energy
information from high energy uncertainties it is necessary that SUSY
be realized at
low energies near the weak scale. In particular there have to be
SUSY partners for the ordinary particles, called superparticles
or sparticles, with masses around the weak scale. The existence of
these
sparticles around the weak scale has significant consequences for
the way coupling constants evolve as you go from low energies to high
energies. As the coupling constant evolves, every time it encounters
a
superparticle the theory becomes less asymptotically free and
therefore
the coupling constant starts evolving more slowly. Therefore a
generic
feature of SUSY theories is that coupling constants, as you
go from low to high energies, evolve more slowly which means that, if
they
have any tendency to meet they meet later than they would have met in
a
non-SUSY theory.
\subsection{Coupling Constant Unification}
This means that if the coupling constants are going to come together
at all
they are going to do so at a point which is later than a non-SUSY
theory.
This in turn implies that the fundamental scale at which coupling
constants
get unified is bigger so that the proton decay rate is slowed down.
For
similar reasons, having to do with the superparticle spectrum, the
weak
mixing angle changes.

In GUTs in general, and in SUSY ones in particular,
the low energy coupling constants $\alpha_1\;\alpha_2$ and $\alpha_3$
are
given in terms of just two fundamental parameters at high energies,
namely
the common coupling constant at the unification mass and the
magnitude of
the unification mass. Since three low energy parameters are given in
terms
of two, there is one prediction, which can be expressed in many ways.
One
possible fruitful way to express it is as a relation between $\sin^2
\theta_W$ and $\alpha_s$ at the mass of the $Z^0$ -- that is, at low
energies.

This relation was worked out many years ago both for non-SUSY
\cite{helen}
and SUSY theories [4,5], see figure 1.
\widefigure{110mm}{The correlation in the values of $\sin^2\theta_W$
and
$\alpha_s(M_Z)$ predicted in SUSY GUTs and ordinary GUTs. The bare
superstring prediction is the point on the far right. The present
1994 are
contrasted with the 1981 data. The bands are the uncertainties in the
theoretical predictions of GUTs and SUSY GUTs. The numbers in the
bands
indicate the unification scale. The uncertainties in the theoretical
predictions for superstrings are not known.}
The data point is for the present measurement.
The numbers shown (15, 16, 17 {\it etc.}) correspond to the logarithm
of
the energy at which unification occurs so that unification in SUSY
theories occurs at $\approx 2\times 10^{16}$ GeV. It can clearly be
seen
that the non-SUSY SM is excluded in view of the recent data
relative to the SUSY SM. It may also be noted that in the
non-SUSY model the unification mass is relatively
small, around $10^{13}-10^{14}$ GeV.

In 1981 a couple of conclusions were drawn: first, the value of
$\sin^2
\theta_W$ for SUSY GUTs is bigger than for the non-SUSY theories; and
second, because of the large magnitude of the unification mass, the
proton
is stable in paractice. It is interesting to recall the state of
experimental
affairs back in 1981. Just around the time this theory was
constructed and
reported in the Second Workshop on Grand Unification in Michigan
(April,
1981) \cite{mich} there were reports of measurements of
$\sin^2\theta_W$
and $\alpha_s$. Of course the error bars were bigger but the central
values of both $\sin^2\theta_W$ and $\alpha_s$ were in closer
agreement
with the non-SUSY case than the SUSY one. This was a
very strong motivation for pursuing proton decay experiments; already
at
that conference candidate proton decay events were repoprted
\cite{pdec}.

The following quote is from Marciano and Sirlin \cite{quote} and
reflects
the prevailing attitude about non-SUSY grand unification in April
1981.
\begin{quote}
The basic idea of grand unification is very appealing. The simplest
model
based on $SU(5)$ has scored an important success in predicting a
value of
$\sin^2\hat{\theta}_W(M_W)$ which is in excellent agreement with
recent
experimental findings (after radiative corrections are included). It
makes
an additional dramatic prediction that the proton will decay with a
lifetime
in the range $10^{30}\sim 10^{32}$ years. If correct, such decays
will be
seen by the planned experiments within the coming year ({\it or may
already
have been seen})\footnote{My italics.}. An incredible discovery may
be
awaiting us.
\end{quote}
So in the beginning SUSY unification appeared to be dead even before
it
started; nevertheless as you know the data evolved. The fact that the
discrepancy resolved itself in favour of the SUSY theory added an
element
of surprise to the history of SUSY unification and perhaps accounts
in part for the great popularity of these ideas today.
\subsection{Precision electroweak measurements}
There are many tests that a theory must pass. One of the subjects
that I
will only briefly discuss is how SUSY does on precision electroweak
data
in terms of the well known $\epsilon_{1,2,3}$ parameters [21,22].
Roughly
speaking $\epsilon_1$ measures the breaking of up--down symmetry and
$\epsilon_3$ measures the breaking of $SU(2)\otimes U(1)$ or the
number of
$SU(2)\otimes U(1)$ breaking mass terms in the theory. Figure 2 shows
the
experimental data from the LEP Collaborations with and without the
SLD
data.
\Figure{65mm}{The $1\sigma$ error eclipses for $\epsilon_1$ and
$\epsilon_3$
using only LEP and LEP plus SLC data. Also shown is the standard
model
prediction for different values of top and Higgs masses. This figure
is
taken from Altarelli [22].}
The standard model gives a beautiful fit for a top mass of 170
GeV$/c^2$.

What happens when you add SUSY? Well SUSY of course has
extra parameters which determine the masses of sparticles and each
dot in
figure 3 represent a different choice of SUSY parameters.
\Figure{85mm}{The values of $\epsilon_1$ and $\epsilon_3$ in SUSY for
a set
of parameter values, the top mass is fixed at 174 GeV$/c^2$. This
figure is
taken from Altarelli [22].}
As can be seen a class of SUSY theories lies within the preferred
ellipse.
Note that the ellipse is a 39\% probability ellipse so that it is not
the
end of the world if you are not exactly within it; plenty of SUSY
theories
are within one or two standard deviations. The moral is fairly
straightforward: as long as you do not have extremely light
sparticles SUSY
can easily be consistent with the high precision electroweak data.
\subsection{Flavour Changing Neutral Currents}
The same holds true for flavour changing neutral currents, which
typically
place an extremely strong constraint on theories. This was an early
difficulty for technicolor theories, wheras in SUSY it is possible to
avoid
it by having degeneracy between squarks as was postulated in the
first
SUSY GUT.
\subsection{Sin $^2\theta_W$ Predictions}
The next question I would like to turn to is, how unique is SUSY
in making these predictions? In table 1 I try to compare non-SUSY
unified theories with SUSY unified theories and strings that do
not have unification below the string scale -- truly single scale
string
theories.
\begin{table*}
\Table{|c|c|c|c|c|}{
& Experiment & $SU(5)$ & SUSY $SU(5)$ & Bare \\
&&&& Strings \\
\hline
$\alpha_s(M_Z)$ & $0.118\pm0.007$ & 0.07 & $0.125\pm.010$ & $0.20+$?
\\
\cline{3-5}
& & $7\sigma$ & okay & $11\sigma+$? \\
\hline
$\sin^2\theta_W(M_Z)$ & $.2317\pm0.0004$ & 0.2141 & $0.2330\pm.0025$
&
$0.221+$? \\
\cline{3-5}
& & $44\sigma$ & okay & $26\sigma+$? \\ }
\caption{The experimental values for $\sin^2\theta_W$ and
$\alpha_s(M_Z)$
are contrasted with the predictions of three theories: ordinary GUTs,
SUSY
GUTs and bare superstrings. Under each prediction we list the number
of
standard deviations that it differs from experiment. GUTs and SUSY
GUTs
predict one of either $\sin^2\theta_W$ and $\alpha_s(M_Z)$; the other
one
is an input. For stings both $\sin^2\theta_W$ and $\alpha_s(M_Z)$ are
predictions. The uncertainties in the theoretical predictions for
superstrings
are not known. }
\end{table*}
In GUTs, whether SUSY or not, you do not predict both $\alpha_s$
and $\sin^2\theta_W$, instead given one you predict the other.
Looking at
the number of standard deviations theory is from experiment, we see
that in
non-SUSY $SU(5)$ if you fit $\sin^2\theta_W$ and predict $\alpha_s$
it is off by quite a bit, but more to the point it gives a very low
unification mass, just $8\times 10^{13}$ GeV$/c^2$, so that the
proton would
decay at a very rapid rate. Similarly if you take $\alpha_s$ from
experiment
and predict $\sin^2\theta_W$ you are off by quite a bit, and again
you get a
low unification mass $3\times 10^{14}$ GeV$/c^2$. SUSY GUTs work
well,
within one or two standard deviations, as you can see from the
numbers.

The predictions that we quote for superstrings assume the minimal
supersymmetric particle content up to the string scale $M_s$ of about
$4\times10^{17}$ Gev$/c^2$ and do not include any potentially large
string induced corrections\footnote{Since the string scale is 20
times the
SUSY GUT scale, the prediction for the proton mass is 20 GeV$/c^2$.}.
These corrections are model dependent: in the absense of a model, it
is not
possible to estimate their magnitude. It is clear that the
corrections
would have to be quite large to make up for the large discrepancies
with
experiment. It is possible that a model will be found where the
corrections
are large and can be tuned to accomodate the data. Such a ``fix"
would
be no better than accomodating ordinary $SU(5)$ with large
corrections
caused by random unobserved multiplets. Also to quote Barbieri {\it
et al.}
\cite{1gp},
\begin{quote}
Why should these corrections maintain the relations between the
couplings
characteristic of the grand unified symmetry, if such a symmmetry is
not
actually realized.
\end{quote}
A simple possibility is that at $M_s$ the string theory breaks to a
SUSY GUT [23,24]; this is a promising new direction which
may combine some of the virtues of both SUSY GUTs and strings. A
challenge of such attempts would be to explain the ratio of the SUSY
GUT
to the string scale.
\section{THE TOP QUARK}
Finally the top quark has been announced \cite{disct}. It is the only
quark
that has a reasonable
mass; you don't need any small parameter to understand its mass and
at first sight if you did not know anything about the world you would
have
guessed that all of the quarks would have the same mass.
\begin{equation}
m_t\approx
v\sim\sqrt{G_F}\hspace{5mm}\Longrightarrow\hspace{5mm}\lambda_t
\approx 1
\end{equation}
Of course this is not the case, other quarks have much smaller masses
which means that they must have much smaller coupling constants,
$\lambda
\ll 1$, which requires symmetries. What I will now try
to discuss is how nicely the top quark fits into SUSY; this is a
qualitative virtue, not a quantitative virtue like the weak mixing
angle.
\subsection{Infrared Fixed points}
A most interesting idea about computing the top quark mass first in
the
context of non-SUSY theories was discussed back around 1980 by
Pendleton and Ross \cite{pross} and by Hill \cite{hill}. They
point out that if the top Yukawa coupling (or top mass) is large
enough,
not too much smaller than unity, then a broad range of initial
conditions
will give rise to the same top quark mass, or Yukawa coupling, at low
energy; see figure 4.
\begin{figure}
\caption{The top quark Yukawa coupling as a function of energy in the
SM.}
\end{figure}
This idea makes a prediction about the top quark: if you follow
Pendleton
and Ross this gives a top quark mass around 240-250 GeV$/c^2$ in the
SM,
which is fairly insensitive to initial conditions, and as a
by-product you
also obtain an upper limit on the top quark mass. The reason for this
behavior is a classic fixed point behavior in the equations which
determine
the evolution of the top quark and strong coupling.
\begin{equation}
16\pi^2\frac{d\lambda_t}{dt}=\lambda_t(C_0\lambda^2_t-C_3g^2_3)\hspace{5
mm}
16\pi^2\frac{dg_3}{dt}=-b_3g^3_3
\end{equation}
where $C_0,\;C_3$ and $b_3$ are constants whose values depend on the
theories particle content.
\begin{equation}
\Longrightarrow
16\pi^2\frac{d}{dt}\left(\frac{\lambda_t}{g_3}\right)=C_0
\left[\lambda^2_t-\frac{(C_3-b_3)}{C_0}g^2_3\right]
\end{equation}
This equation shows that $\lambda_t$ tracks $g_3$: if the top quark
is too
heavy then the Renormalization Group Equations (RGEs) push it down or
if it
is too small they increase it. This is a stable fixed point:
\begin{equation}
\left(\frac{\lambda_t}{g_3}\right)^2\approx \frac{(C_3-b_3)}{C_0}
\hspace{5mm}\Longrightarrow\hspace{5mm}\frac{d}{dt}
\left(\frac{\lambda_t}{g_3}\right)\approx 0
\end{equation}
accounting for the behavior described above.

This is a nice idea which has been generalized to SUSY \cite{gens}
and leads
to interesting results. Of course in SUSY there are two Higgs fields,
so what
you obtain as a result of doing the same analysis is not exactly an
absolute
mass for the top quark but a scale for the top quark times the sine
of an
angle that measures the ratio of the vacuum expectation values of the
two
Higgses. \begin{equation}
m_{top}=190\;\mbox{GeV}/c^2\;\times\sin\beta
\end{equation}
So the experimental range for $m_t$ is consistent with the top being
near its
SUSY fixed point. This is helpful for bottom--tau unification
\cite{chan}.
That is, if you want to have the bottom and the tau masses equal at
the
grand scale, which occurs in many GUTs, then you have to be within
10\% of
the fixed point for the top quark mass \cite{btau}. So there is a
nice
connection between being at the fixed point and other ideas.
\subsection{Upper Bounds on the Lightest Higgs Mass}
Another thing which you gain by being at the fixed point is that you
improve
a great deal the upper limits on the Higgs mass that you have in
SUSY theories; see figure 5.
\Figure{50mm}{The correlation between the top and lightest
Higgs mass for $\alpha_s=0.110-0.125$ (region between the dotted
lines).
Also shown is the upper bound on the Higgs mass in the minimal SUSY
SM. This figure is taken from Barbieri {\it et al.} [30].}
The solid line is the upper limit for the Higgs mass and is fairly
model
independent \cite{hemp}. It depends
logarithmically on the mass of the stop; as long as the stop is not
much
heavier than 1 TeV$/c^2$ this is a good upper limit. Now if you
assume
that you are near the top quark fixed point you obtain a much
stronger
upper limit given by the two dotted lines, the range depends on
details
like the precise value of $\alpha_s$, but you can see for example
that
for $m_t=170$ GeV$/c^2$ the upper limit goes down to about 90
GeV$/c^2$ from
about 150 GeV$^2$. So if you are near the top fixed point this
significantly
pushes down the upper limit to the lightest Higgs mass.
\subsection{Superparticle Spectra}
Another virtue of being the top quark fixed point is that you have a
reduction in the number of parameters that determine the spectrum of
the
superparticles. Instead of the usual set of parameters
\begin{equation}
m_{1/2},\;\;m_0,\;\;B,\;\;\mu,\;\;\lambda_t,\;\;A
\end{equation}
you can compute the full superparticle spectrum in terms of two
parameters
and there are also some simplifications which emerge. For example the
gauginos become $SU(2)\otimes U(1)$ eigenstates and you get simple
mass
relations
\begin{eqnarray}
M_{\tilde{W}^\pm}&=&M_{\tilde{W}^0}=\frac{\alpha_2}{\alpha_1}M_{\tilde{B
}}
=2M_{\tilde{B}} \\
M_{\tilde{g}}&\approx&\frac{\alpha_3}{\alpha_1}M_{\tilde{B}}
\end{eqnarray}
and in general the spectrum of superparticles becomes much more
manageable
[17,32].
{\it 3.4~Dynamical Determination of the Top Yukawa Coupling}
An interesting related idea is that of Kounnas, Zwirner and Pavel
\cite{koun}
and
Binetruy, Dudas and Pillon \cite{bine} who give dynamical reasons why
you might
be near the top quark fixed point. They argue that if there is a
field that
slides, for example a modulus field, on which the top Yukawa coupling
depends
and this is the only place where this field appears then minimizing
the
effective potential of that field:
\begin{equation}
\frac{\partial V_{eff}(\varphi)}{\partial\varphi}=0=
\frac{\partial V}{\partial\lambda_t}
\frac{\partial \lambda_t}{\partial\lambda^{GUT}_t}=0
\end{equation}
has a solution that corresponds to the fixed point.
\begin{equation}
\frac{\partial\lambda_t}{\partial\lambda^{GUT}_t}=0
\end{equation}
A solution which says that the low energy top Yukawa coupling is
insensitive
to its grand unified value. They also argue that this fixed point
solution
may be the lowest minimum of this potential and this may give, in
such
theories, a dynamical reason for being near the fixed point. These
authors
are pursuing these ideas further and are trying to argue about the
smallness
of the bottom with respect to the top Yukawa coupling to explain the
lightness
of the bottom quark mass.
\subsection{Vacuum Stability in the Standard Model}
So far we have shown that in SUSY theories we get an upper bound on
the lightest Higgs, what happens in the non-SUSY theories -- in
the SM? Actually in the SM you get a lower bound on the mass of the
Higgs
and this happens because of vacuum stability. If $\lambda$ is the
quartic
coupling of the Higgs field which is responsible for the mass of the
Higgs
then of course $\lambda$ has to be positive to have a stable
Hamiltonian
which is not unbounded from below [34,35]. However if you have a
large
top Yukawa coupling the RGEs for the Higgs mass (or the quartic
coupling)
have a positive term and a negative term where the negative term
depends on
the fourth power of the top Yukawa coupling.
\begin{equation}
\frac{d\lambda}{dt}=\lambda^2-\lambda^4_t
\end{equation}
So if the top Yukawa coupling is large, which of course it must be
for the
top to be heavy, then  you are potentially driven to a negative
$\lambda$
and an unstable situation. To prevent this instability you have to
have a
large quartic coupling which means a lower limit on the Higgs mass.
This
lower limit can be computed and is shown in figure 6.
\Figure{65mm}{The lower limit on the Higgs mass as a function
of the unification scale $\Lambda$ for $m_t=174$ GeV$/c^2$. This
figure is
taken from Altarelli {\it et al.} [35].}
It is not very sensitive to the top mass. The scale $\Lambda$ in the
standard, non-SUSY model is the scale at which new physics must
enter, namely the scale at which the vacuum destabilizes. So if you
want
to have a stable vacuum up to the GUT mass or up to the Planck mass
there
is a lower limit to the higgs mass ${\cal O}(135)$ GeV$/c^2$. To
summarize
in the SM you get a lower limit to the Higgs mass if the SM is valid
all
the way up to the GUT scale and in the SUSY SM you get an upper
limit.
\section{OTHER QUARK AND LEPTON MASSES}
So far we have focussed on just two parameters the weak mixing angle
and
the top mass. Of course the theory has many more parameters: the
SUSY theory has another 20 parameters about which have not said
anything. 14 of these parameters have to do with the quark masses and
mixing angles. When theorists try to attack this problem they have to
confront a big disadvantage relative to the experimentalists.
The experimentalist have only 14 (6+3+3+1+1) parameters to measure,
the
theorists {\it a priori} has three $3\times3$ matrices each element
of
which is a complex number. Therefore the theorist starts out with 54
(=
$3\times3\times3\times3\times2$) parameters and wants to explain some
of
these fourteen.

The idea of grand unification is a great help in reducing the number
of
parameters. GUTs can relate the lepton masses to the down and up
masses, as
well as neutrino masses, and in $SO(10)$ GUTs all of these mass
matrices are
related to each other. Therefore it suffices to focus on one of these
mass
matrices, say the electron or negatively charged lepton mass matrix
to
explain the rest. Now the number of free parameters has been reduced
to one
$3\times3$ matrix which after removing some phases leaves 16
parameters.
Since 16 is larger than 14 so you have still not quite begun to
predict
something of relevance for experiment.

In order to ameliorate this situation it is clear what we have to do.
Grand unified gauge symmetry is not sufficient to make predictions
you
need some more symmetry you need some flavour symmetry that will
relate
quarks of different families to each other and can perhaps explain
why all
quarks and leptons are not degenerate with the top quark right at the
weak
scale.
\subsection{The Textural Approach}
There are at least two approaches to the problem of fermion masses
and
several people have done interesting work on this. First there is
what
is called the textural approach. Texture refers to the
following: you start with every mass matrix, quark or lepton, as a
$3\times3$ matrix. Since you have to make assumptions to reduce the
number of parameters you can assume a specific pattern of zeros and
symmetry or antisymmetry of this matrix to reduce the number of
parameters.
Some people object that postulating a number to be zero is choosing
it to be a very specific value. The great thing about zero is that
zero
can be the consequence of a symmetry. We can think of
many ways to make something zero; that is the motivation for
choosing zeros as opposed to any other number for specific entries.
There
are also regularities like nearest neighbour mixing {\it etc.} in the
pattern of observed mixing angles that also phenomenologically
motivate
some zeros. This approach was pioneered by Fritsch \cite{frit} and in
the context of GUTs  by Georgi and Jarlskog \cite{geor}. A
lot of work which has been done in the last few years along
these lines \cite{lot}.
\subsection{The Operator Approach}
Then there is a more ambitious approach which you may call the
operator
approach. According to this approach you start with first of all a
SUSY $SO(10)$ theory: SUSY to explain the weak mixing
angle and $SO(10)$ to be able to relate all quarks and leptons to
each
other. You right down the smallest set of operators that you can that
will give masses to all the quarks and leptons. This approach has
been
pursued recently \cite{ops} and it has some quantitative virtues. Its
biggest virtue is that it is very predictive with 6 inputs it can get
14
outputs, namely the parameters of the quark and lepton mass matrix,
thereby
making 8 predictions.

I do not have space to discuss the whole technology that is involved
in
this approach, it is a very technical subject. There is a discrete
scanning procedure that gives you a discrete set of theories of which
three or four survive this test. Table 2 shows an example of the
type of inputs and outputs that are obtained.
\begin{table}
\Table{|c|c|c|c|}{
Input & Input & Predicted & Predicted \\
Quantity & Value & Quantity & Value \\
\hline
$m_b(m_b)$ & 4.35 GeV & $m_t$ & 176 GeV \\
$m_\tau(m_tau)$ & 1.777 GeV & $\tan\beta$ & 55 \\
\hline
$m_c(m_c)$ & 1.22 GeV & $V_{cb}$ & .048 \\
\hline
$m_\mu$ & 105.6 GeV & $V_{ub}/V_{cb}$ & .059 \\
$m_e$ & .522 MeV & $m_s$(1 GeV) & 172 MeV \\
$V_{us}$ & .221 & $\hat{B}_K$ & .64 \\
&& $m_u/m_d$ & .64 \\
&& $m_s/m_d$ & 24. \\ }
\caption{Predictions of a class of models from [39].}
\end{table}
If you input the six number in the first column, which are very well
known,
you output 8 numbers. In addition you predict very precisely things
having
to do with CP violation \cite{jarl}:
\begin{eqnarray}
\sin2\alpha & = & -.46 \\
\sin2\beta & = & -.49  \\
\sin2\gamma & = & -.84 \\
J & = & + 2.6\times 10^{-5}
\end{eqnarray}
These predictions are sufficiently sharp that they can be tested in
the
B- factory for example.
\section{CONCLUSIONS}
To conclude I would like to state some of my personal biases which
are
actually shared by many people. These biases are of course not time
independent\footnote{As of 4:04 p.m. July 25th, 1994.}

The first bias is a quantitative one namely that SUSY GUTs are
correct on the basis of the evidence we have for the weak mixing
angle.
As I have said this could be a 2 in 100 accident and in that case we
have
nothing to go by. Then there are some qualitative virtues of SUSY
GUTs, of these I have discussed:
\begin{itemize}
\item Naturalness: the light sector decouples from the heavy sectors
of the
theory. The decoupling is not total the weak mixing angle still
depends
logarithmically on $M_{PL}$.
\item The non-observation of proton decay; this is only a qualitative
virtue since any theory which does not unify shares it.
\item The fact that the top quark fixed point fits nicely within
SUSY.
\end{itemize}
Also other virtues that are consistent with having a heavy top quark,
that I
did not have space to discuss, also fit nicely in the context of
SUSY.
\begin{itemize}
\item Bottom--tau unification.
\item Radiative electroweak symmetry breaking \cite{radb}.
\end{itemize}

Of course the big question is how shall we know if SUSY is really
there? and when. The easy answer is when LHC and NLC are built. In
order to
have decoupling of the weak world from the Planckian world we need
SUSY
particles to exist around a TeV or below. Or course this is not a
hard number
it is an estimate. The first consequence of SUSY is that all
sparticle masses are roughly less than 1 Tev. Before LHC and NLC
there is
still hope that we may see something for example proton decay.
\vspace{5mm} \\
\noindent
{\bf Proton Decay} [17,42] SUSY predicts that the proton can decay at
a
reduced rate into kaons.
\begin{equation}
p\rightarrow K^++\bar\nu\hspace{10mm}n\rightarrow K^0+\bar\nu
\end{equation}
These are very unique modes; they are general consequences of SUSY
theories under very general conditions. The strongest ingredient is
Fermi
statistics so it is not a highly model dependent
statement that nucleons decay into kaons. Icarus and Superkamiokande
may
get lucky and with limits ${\cal O}(10^{34})$ years may be able to
see such
events. I should say that in contrast to non-SUSY theories
SUSY does not make a sharp prediction about the proton lifetime
because proton decay is mediated by very heavy Higgs-like particles
(not gauge
particles) whose coupling constants are not very well under control,
so
this is not a hard prediction.
\vspace{5mm} \\
\noindent
{\bf Neutron and electron electric dipole moments} (edm). If you take
a SUSY
theory with sparticles
around 100 GeV$/c^2$ and phases of order unity you find that the edm
of
the neutron is $10^{-23}$ ecm which is a factor of 100 too large.
This
is not a deadly diseases because we do not know the masses of
sparticle or
their phases. However it suggests that if the limits for edms improve
by
a factor of 10 or 100 then there is a good chance if SUSY is right
that something may be seen and actually if nothing is seen it is
reason
to start wondering about SUSY.
\vspace{5mm} \\
\noindent
{\bf Flavour surprises} There are many possibilities for these
because to
ensure that there are no FCNC you have to assume degeneracy in
sparticle
masses which is broken by weak effects. Therefore flavour surprises
in SUSY theories are possible and the B-factory, for example,
may be a place to look for these things or for anything that has to
do with
theories that predict CKM elements and fermion masses.
\vspace{5mm} \\
\noindent
{\bf Neutrino masses} I really do not have any idea what to say about
neutrino
masses. To make any statement you have to make a long list of
assumptions
that one does not have very strong faith in. Chorus, Nomad and
hopefully
the long baseline experiments will be able to resolve this. Lots of
SUSY GUTs have:
\begin{equation}
\mu\rightarrow e\gamma
\end{equation}
\vspace{5mm} \\
The only hint for perhaps some physics beyond the SM is the weak
mixing
angle. To predict it you need to simultaneously postulate unification
of
the couplings constants with an $SU(3)\otimes SU(2)\otimes U(1)$
desert
and low energy SUSY, namely particles at accessible energies.
The weak mixing angle depends on
the integrated effects of virtual SUSY that extend from the
Planck mass all the way down to the weak scale. So if SUSY turns
out to be right it will be fascinating that the virtual effects of
the superparticles that propagate information down from the Planck
mass
to the weak scale will have been seen before the actual live
superparticles
themselves.
\noindent
\section*{Acknowledgements}
It is a pleasure to thank G.\ Altarelli, R.\ Barbieri, M.\ Carena,
N.\ Polonsky and R.\ Wagner for many valuable conversations. I would
also like to thank I.\ Knowles for preparing this document from the
video recording of my talk and for his valuable suggestions.
{\bf 	QUESTIONS}
{\it A.V.\ Efremov, JINR--Dubna:} \\
In Vysotsky's talk (these proceedings) we heard about one more
indication
of supersymmetry connected with $b$-quark decay. Do you consider it
to be
important also?
\vspace{5mm}\\
{\it S.\ Dimopoulos:} \\
I think it is an interesting result. I would not go so far as to call
it
another indication for SUSY; not yet.
\vspace{5mm} \\
{\it J.L.\ Chkareuli, IoP--Tiblisi/Sussex:} \\
I have a little comment concerning the unification of the standard
coupling constants. Actually, it is not a privilege of the SUSY
$SU(5)$
model only. We found many Extended GUTs from $SU(6)$ giving a perfect
unification in the non-SUSY case. All these EGUTs containing a number
of additional pairs of conjugated multiplets in their fermion
spectrum
are proved to be broken not through the standard $SU(5)$ but through
alternative channels. Besides the minimal $SU(6)$ case there are good
examples of natural unification in $SU(9)$ (Frampton) and $SU(11)$
(Georgi) models including the gauged quark lepton families.
\vspace{5mm} \\
{\it S.\ Dimopoulos:} \\
you are refering to theories with intermediate scales.
If you have intermediate scales you do not predict $\sin^2\theta_W$,
$\sin^2\theta_W$ is an input and then you predict some phenomena at
some scale ${\cal O}(10^{10})$ GeV. So this is not an experimentally
testable success.
\vspace{5mm} \\
{\it G.G.\ Ross, Oxford:} \\
I should like to point out that $M_{string}$ is not the unification
scale
in an arbitrary string theory -- it must be determined for the
specific
string theory. In the absence of this information you must use $M_X$
as a
free parameter -- just as in SUSY GUTS.
\vspace{5mm} \\
{\it S.\ Dimopoulos:} \\
Indeed if you succeed in constructing a string theory that breaks at
$M_s=4\times10^{17}$ GeV$/c^2$ down to a SUSY GUT (with the usual
SUSY
GUT scale of $M_X=2\times10^{16}$ GeV$/c^2$)  you will have succeeded
in combining the virtues of SUSY GUTs with those of strings. This is
precisely the program of Ibanez {\it et al.} and Lykken {\it et al.}
that I refered to in my talk. As I explained in my talk the
predictions
that I quoted for ``bare" superstrings are what you get if you the
minimal supersymmetric particle content up to $M_s$, no intervening
unification and no large threshold corrections to fix things up.
\vspace{5mm}\\
{\it G.\ Crosetti, INFN--Genova:} \\
How robust is the upper limit on the Higgs masses? Because if it is
really very strong LEP-II can state something on the SUSY model in
the
next few years. Do you agree with this?
\vspace{5mm}\\
{\it S.\ Dimopoulos:} \\
It depends logarithmically on the assumed sparticle masses.
\vspace{5mm}\\
{\it H.\ Haber, UCSC} \\
It also relies on you having $\tan\beta$ rather small which is why
the
limit is so strong.
\vspace{5mm}\\
{\it S.\ Dimopoulos:} \\
I recall the upper limit to be 160 GeV$/c^2$ (even if sparticles are
at 10
TeV$/c^2$) for any value of $\tan\beta$ up to $\sim 60$. On my
transparency
I showed what happens up to $\tan\beta=10$ because the upper limit
does
not change much for $\tan\beta>10$
\vspace{5mm}\\
{\it H.B. Nielson, NBI} \\
I would like to mention our work on trying to predict the fine
structure
constants.
\vspace{5mm}\\
{S.\ Dimopoulos:} \\
If I recall correctly your predictions have an uncertainty of
$\sim\,\pm
20$\%. The experimental accuracy on $\sin^2\theta_W$ is $\pm0.2$\%.
Ordinary GUTs are off by only $\sim 10$\% on $\sin^2\theta_W$, yet
this
means 40 standard deviations. It is hard to draw a conclusion until
you
improve the accuracy of your calculations.

\begin{thebibliography}{99}
\bibitem{unif}
H.\ Georgi and S.\ Glashow, \prl{32}{74}{438}; \\
J.C.\ Pati and A.\ Salam, \prev{D8}{73}{1240}.
\bibitem{helen}
H.\ Georgi, H.\ Quinn and S.\ Weinberg, \prl{33}{74}{451}.
\bibitem{tech}
S.\ Weinberg, \prev{D13}{76}{974}\ and\ \ib{D19}{79}{1277}; \\
L.\ Susskind, \prev{D20}{79}{2619}; \\
E.\ Fahri and L.\ Susskind, \prep{74}{81}{277}.
\bibitem{sunf}
S.\ Dimopoulos and H.\ Georgi, in the 2nd Workshop on Grand
Unification, Michigan 1981, p.285; Eds. J.P.\ Leville, L.R.\ Sulak
and D.C.\ Unger (Birkh\"{a}user 1981); \\
S.\ Dimopoulos and H.\ Georgi, \np{B193}{81}{150}.
\bibitem{sunif}
S.\ Dimopoulos, S.\ Raby and F.\ Wilczek, \prev{D24}{81}{1681}.
\bibitem{string}
See, M.B.\ Green, J.H.\ Schwarz and E.\ Witten, Superstring Theory
(Cambridge 1987) and references therein; \\
M.B.\ Green, these proceedings.
\bibitem{discw}
UA2 Collaboration: Banner {\it et al.}, \pl{122B}{83}{476}.
\bibitem{discz}
UA1 Collaboration: G.\ arnison {\it et al.}, \pl{126B}{83}{398}; \\
UA2 Collaboration: Banner {\it et al.}, \pl{129B}{83}{130}.
\bibitem{disct}
CDF Collaboration: Abe {\it et al.}, \prl{}{94}{}. \\
H.\ Jensen, these proceedings; \\
P.\ Granis, these proceedings.
\bibitem{prec}
D.\ Schaile, these proceedings; \\
M.\ Vysotsky, these proceedings.
\bibitem{hint1}
K.\ M\"{o}nig, these proceedings.
\bibitem{hint2}
M.J.\ Fero, these proceedings.
\bibitem{subs}
See the listing these proceedings.
\bibitem{bull}
The Los Alomos National Laboratory phenomenology archieve:
{\tt hep-ph@xxx.lanl.gov}.
\bibitem{lane}
K.\ Lane, these proceedings.
\bibitem{proc}
See also in these proceedings: \\
S.T.\ Love; \\
J.W.F.\ Valle; \\
H.B\ Nielson, D.L.\ Bennett and C.D.\ Froggatt; \\
M.J.\ Duff; \\
S.F.\ King, T.\ Elliot and P.L.\ White; \\
J.F.\ Gunion \\
C.D.\ Froggatt \\
G.\ Zoupanos, J.\ Kubo and M.\ Mondragon; \\
E.\ Dudas; \\
M.\ Matsuda, T.\ Hayashi, Y.\ Koide and M.\ Tanimoto; \\
J.L.\ Chkareuli, I.G.\ Gogoladze and A.B.\ Kobakhidze.
\bibitem{lazy}
V. Barger, M.S.\ Berger and P.\ Ohmann, these proceedings; \\
P. Nath and R. Arnowitt, these proceedings.
\bibitem{mich}
The 2nd Workshop on Grand Unification, Michigan 1981; Eds. J.P.\
Leville, L.R.\ Sulak and D.C.\ Unger (Birkh\"{a}user 1981).
\bibitem{pdec}
Kolar gold field experiment: S.\ Miyake and V.S. Narashimham {\it et
al.},
in \cite{mich} p.11; \\
Homestake mine experiment: R.I. Steinberg {\it et al.}, in
\cite{mich} p.22.
\bibitem{quote}
W.J.\ Marciano and A.\ Sirlin, in \cite{mich} p.160.
\bibitem{uts}
B.W.\ Lynn, M.E.\ Peskin and R.G. Stuart, LEP study group
CERN yellow report 85;\\
M.E.\ Peskin and T.\ Takauchi, \prl{65}{90}{964}\ and\
\prev{D46}{91}{381}; \\
G.\ Altarelli and R.\ Barbieri, \pl{B253}{90}{161}; \\
G.\ Altarelli, R.\ Barbieri and S.\ Jadach, \np{B369}{92}{3}; \\
G.\ Altarelli, R.\ Barbieri and F.\ Caravaglios, \pl{B314}{93}{357}.
\bibitem{eps}
G.\ Altarelli, preprint: CERN-TH-7464/94.
\bibitem{1gp}
R.\ Barbieri, G.\ Dvali and A.\ Strumia, Pisa preprint:
IFUP-PTH-94-22.
\bibitem{3gps}
G.\ Aldazabal, A.\ Font, L.E.\ Ibanez and A.M.\ Uranga, Madrid
preprint:
FTUAM-94-28; \\
S.\ Chadhouri, S.-W.\ Chung and J.D.\ Lykken, Fermilab-PUB-94-137-T;
\\
G.B.\ Cleaver, Ohio preprint: OHSTPY-HEP-T-94-007.
\bibitem{pross}
B.\ Pendleton and G.G.\ Ross, \pl{98B}{81}{291}.
\bibitem{hill}
C.T.\ Hill, \prev{D24}{81}{691}.
\bibitem{gens}
L.\ Alvarez-Gaume, J.\ Polchinski and M.B.\ Wise, \np{B221}{83}{495};
\\
L.E.\ Ibanez, and C.\ Lopez, \pl{126B}{83}{54}; \\
J.\ Bagger, S.\ Dimopoulos and E.\ Masso, \pl{156B}{85}{357}.
\bibitem{chan}
M.S.\ Chanowitz, J.\ Ellis and M.K.\ Gaillard, \np{B128}{77}{506}; \\
A.J.\ Buras, J.\ Ellis, M.K.\ Gaillard and D.V.\ Nanopoulos,
\np{B135}{78}{66}; \\
L.E.\ Ibanez and C.\ Lopez, \pl{126B}{83}{54}.
\bibitem{btau}
A.\ Giveon, L.J.\ Hall and U.\ Sarid, \pl{B271}{91}{138}; \\
C.D. Froggatt, R.G. Moorhouse and I.G. Knowles, \pl{B298}{93}{356};
\\
V.\ Barger, M.S.\ Berger and P.\ Ohmann, \prev{D49}{94}{4908}; \\
W.A.\ Bardeen, M.\ Carena, S.\ Pokorski and C.E.M.\ Wagner,
\pl{B320}{94}{110}; \\
M.\ Carena, M.\ Olechowski, S.\ Wagner and C.E.M.\ Carena,
\np{B426}{94}{269}.
\bibitem{hemp}
Y.\ Okada, M.\ Yamaguchi and T. Yanagida, Prog. Theor. Phys. Lett.
{\bf 85}
(1991) 1; \\
J.\ Ellis, G.\ Ridolfi and F. Zwirner, \pl{B257}{91}{83}\ and\
\np{B262}{91}{477}; \\
H.E.\ Haber and R.\ Hempfling, \prl{66}{91}{1815}; \\
R.\ Barbieri, R.\ Frigeni and F.\ Caravaglios, \pl{B258}{91}{67}.
\bibitem{care}
M.\ Carena and C.E.M.\ Wagner, preprint: CERN-TH-7393-94.
\bibitem{koun}
C.\ Kounnas, F.\ Zwirner and I.\ Pavel, \pl{B335}{94}{403}.
\bibitem{bine}
P.\ Binetruy, E.\ Dudas and F.\ Pillon, \np{B415}{94}{175}.
\bibitem{rome}
N.\ Cabibbo, L.\ Maiani, G.\ Parisi and R.\ Petronzio,
\np{B158}{79}{295}; \\
M.\ Sher, \prep{179}{89}{273}.
\bibitem{renew}
G.\ Altarelli and G.\ Isidori, \pl{B337}{94}{141}.
\bibitem{frit}
H.\ Fritsch, \pl{70B}{77}{436};\ and\ \np{B155}{79}{189}.
\bibitem{geor}
H.\ Georgi and C.\ Jarlskog, \pl{86B}{79}{297}.
\bibitem{lot}
J.\ Harvey, P.\ Ramond and D.\ Reiss, \pl{B92}{80}{309};\ and\
\np{B199}{82}{223}; \\
X.G.\ He and W.S.\ Hou, \prev{D41}{90}{1517}; \\
S.\ Dimopoulos, L.J.\ Hall and S. Raby, \prl{68}{92}{752};\ and\
\prev{D45}{92}{4192}; \\
H.\ Arason, D.\ Castano, E.J.\ Pirad and P.\ Ramond,
\prev{D47}{93}{232}; \\
V.\ Barger, M.S.\ Berger, T.\ Han and M.\ Zralek, \prl{68}{92}{3394};
\\
P.\ Ramond, R.G.\ Roberts and G.G.\ Ross, \np{B406}{93}{19}.
\bibitem{ops}
G.\ Anderson, S.\ Dimopoulos, L.J. Hall, S.\ Raby and G.\ Starkman,
\prev{D49}{94}{3660}. \\
For early work in this direction see: \\
C.D. Froggatt and H.B.\ Nielson, \np{B147}{79}{277}; \\
S.\ Dimopoulos, \pl{129B}{83}{417}; \\
J.\ Bagger, S.\ Dimopoulos, H.\ Georgi and S.\ Raby, in Proc. 5th
Workshop
on Grand Unif., Rhode Island 1984; (World scientific 1984).
\bibitem{jarl}
C.\ Jarlskog, \prl{55}{85}{1039}; \\
C.\ Jarlskog and R.\ Stora, \pl{B208}{88}{268}.
\bibitem{radb}
L.E.\ Ibanez and G.G.\ Ross, in {\it Perspectives on Higgs Physics},
p.229;
Ed. G.L.\ kane (World Scientific 1993); and references therin.
\bibitem{pees}
S.\ Dimopoulos, S.\ Raby and F.\ Wilczeck, \pl{112B}{82}{133}; \\
J.\ Ellis, D.V.\ Nanopoulos and S.\ Rudaz, \np{B202}{82}{43}.
\end{thebibliography}
\end{document}